\documentstyle[pra,aps,twocolumn,epsfig]{revtex}

\newcommand{\bra}[1]{\langle #1 | \,}
\newcommand{\ket}[1]{\, | #1 \rangle}

\newcommand{\om}{\omega}
\newcommand{\Om}{\Omega}
\newcommand{\la}{\lambda}
\newcommand{\ga}{\gamma}
\newcommand{\eps}{\epsilon}
\newcommand{\veps}{\varepsilon}
\newcommand{\De}{\Delta}
\newcommand{\de}{\delta}

\begin{document}

\draft

%%% Comment if preprint style is used %%%
\twocolumn[\hsize\textwidth\columnwidth\hsize\csname@twocolumnfalse\endcsname

\title{Photon-photon correlations and entanglement in doped photonic crystals}

\author{David Petrosyan and Gershon Kurizki}

\address{Department of Chemical Physics, Weizmann Institute of Science, 
Rehovot 76100, Israel}

\date{\today}

\maketitle

\begin{abstract}
We consider a photonic crystal (PC) doped with four-level atoms whose 
intermediate transition is coupled near-resonantly with a photonic band-gap 
edge. We show that two photons, each coupled to a different atomic transition
in such atoms, can manifest strong phase or amplitude correlations: One photon 
can induce a large phase shift on the other photon or trigger its absorption 
and thus operate as an ultrasensitive nonlinear photon-switch. These 
features allow the creation of entangled two-photon states and have unique 
advantages over previously considered media: (i) no control lasers are needed; 
(ii) the system parameters can be chosen to cause full two-photon entanglement 
via absorption; (iii) a number of PCs can be combined in a network.
\end{abstract}


%%% Comment if preprint style is used %%%
]

\section{Introduction}

Nonlinear effects whereby one light beam influences another require large 
numbers of photons \cite{nonlin} or else photon confinement in a high-Q cavity
\cite{cavity}. Hence the impediment towards constructing quantum 
logical gates operating at the few-photon level. The ingenious attempt to 
achieve increased photon-photon coupling in a gas by means of control laser  
fields \cite{franson} has resulted in cooperatively enhanced single-photon 
absorption and emission (as for excitons in solids), but not in two-photon 
entanglement \cite{opatrny}. A promising avenue has been opened by studies of 
enhanced nonlinear coupling via electromagnetically induced transparency
(EIT) in gases in the presence of control laser fields, which induce coherence 
between atomic levels \cite{eit}. These studies have predicted the ability 
to achieve an appreciable nonlinear phase shift using extremely weak optical 
fields or a two-photon switch in the $N$-configuration of atomic levels 
\cite{imam,harris}. Further improvement of the sensitivity of these schemes
has been suggested using a rather involved system, in which a second species 
of coherently driven $\Lambda$-atoms provides the matching of the group 
velocities of interacting photons \cite{lukin}.

Here we point out that photon-photon nonlinear phase shifters and switches are
realizable with a potentially very high efficiency and {\it without} external 
laser fields in photonic crystals (PCs) \cite{yablon,john,joannopoulos} 
dilutely doped with four-level atoms. These atoms have two transitions tuned 
to the two incident photons and an intermediate transition tuned to a singular
feature of the structured density of modes (DOM) spectrum of the PC. The strong
nonlinear effects analyzed in Secs. \ref{theory} and \ref{results} arise from 
the coherent Autler-Townes splitting of atomic emission lines near a photonic 
band-gap (PBG) edge \cite{kofman,john_wang,john_quang,pl} and the strong 
photon-atom coupling via photonic defect modes in the doped PC 
\cite{defect,kur}. Transparency near a band edge, which has been previously 
predicted for an atomic three-level $\Lambda$-configuration \cite{knight}, is 
shown here (Sec. \ref{results}) to be most suitable for {\it two-photon 
absorption switching}. This mechanism, which has not been studied in PCs, is 
demonstrated to be considerably more efficient than its counterpart in 
Ref. \cite{harris}. It is predicted to allow {\it complete} absorption of one 
photon in the presence of another photon, and thereby the creation of a 
{\it fully entangled} two-photon state (Sec. \ref{discuss}). By contrast, 
appreciable (but limited) {\it nonlinear phase-shifting} is shown 
(Sec. \ref{results}) to arise by tuning the photon frequency to the Raman 
resonance with the defect mode frequency associated with atomic doping. 
These features have unique advantages for quantum information applications 
(Sec. \ref{discuss}), as compared to previously considered media.

\section{Theory}
\label{theory}

\begin{figure}[htb]
\centerline{\psfig{figure=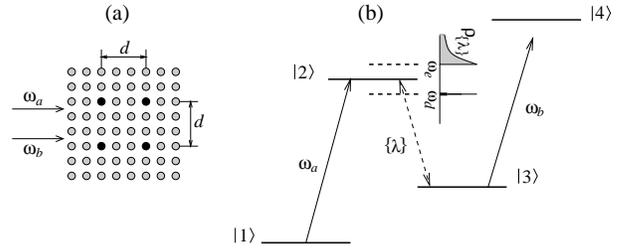,width=8cm}}
\vspace{0.3cm}
\caption{(a) Photonic crystal dilutely doped with atoms located at black
dots. (b) Four-level atom coupled to a structured continuum \{$\la$\} near
the band-edge or defect mode frequencies (DOM plotted) via the intermediate 
transition $\ket{2} \to \ket{3}$ and interacting with two photons $\om_a$ and 
$\om_b$ at the sideband transitions $\ket{1}\to \ket{2}$ and 
$\ket{3}\to \ket{4}$, respectively.}
\label{at_sys}
\end{figure}

We examine the nonlinear coupling of two optical fields  
$\hat{E}_a = f_a \veps_a \hat{a}$ and $\hat{E}_b = f_b \veps_b \hat{b}$, 
where $\hat{a}$ and $\hat{b}$ are the respective annihilation operators, 
$\veps_i= (\hbar \om_i/2 \eps_0 V_i)^{1/2}$ is the field amplitude per photon 
$\om_i$ ($i=a,b$) within the quantization volume $V_i=\sigma_i l_i$, and 
$f_i(z,t)$ is the single-photon wavepacket envelope. The two fields 
propagate along the $z$-axis in a PC dilutely doped with identical four-level 
atoms. The level configuration of the atom and the DOM are depicted in
Fig. \ref{at_sys}, where the unperturbed atomic levels $\ket{j}$, $j=1\dots 4$,
have the corresponding energies $\hbar \om_j$. The incident photons at 
frequencies $\om_a$ and $\om_b$ interact with the atoms via the transitions 
$\ket{1} \to \ket{2}$ and  $\ket{3} \to \ket{4}$, respectively, while the 
transition $\ket{2} \to \ket{3}$ is coupled to the structured PC mode-continuum
\{$\la$\}. In the dipole and rotating-wave approximations, the Hamiltonian of 
the $l$th atom+field can be written as
\begin{eqnarray}
H^{(l)} &=& \sum_j \hbar \om_j \hat{\sigma}_{jj}^{(l)}- \hbar \Bigl[ 
\sum_{\la} g_{\la} \hat{c}_{\la} 
e^{i (k_{\la} z - \om_{\la} t)} \hat{\sigma}_{23}^{(l)}
\nonumber \\ & & 
+ g_a \hat{a} e^{i (k_a z - \om_a t)} \hat{\sigma}_{21}^{(l)} +
g_b \hat{b} e^{i (k_b z - \om_b t)} \hat{\sigma}_{43}^{(l)} 
+ \text{H. c.} \Bigr]  , \label{ham}
\end{eqnarray}
where $k_i$ ($i = a,b,\la$) is the wave number of the corresponding mode, 
$g_i = \mu_{jk} f_i \veps_i /\hbar$ its atom-field coupling strength, 
$\mu_{jk}$ being the dipole matrix element for the atomic transition 
$\ket{j} \to \ket{k}$, $\hat{\sigma}_{jk}^{(l)} = \ket{j} \bra{k}$ is the 
atomic operator, $\hat{c}_{\la}$ is the $\lambda$ mode annihilation operator 
and 
$\sum_{\la} \to \int d \om_{\la} \rho(\om_{\la})$, where $\rho(\om_{\la})$ 
is the DOM of the structured continuum.

We assume that initially the two incident photons are in the product state 
$\ket{1_a} \ket{1_b}$, the atoms are in the ground state $\ket{1}$ and the 
continuum is in the vacuum state $\ket{0_{\la}}$. Then the wave function of 
the system reads 
$\ket{\Phi (z_l,t)}= A_1 \ket{1,0_{\la},1_a,1_b} + A_2\ket{2,0_{\la},0_a,1_b} 
+\sum_{\la} A_{3,\la} \ket{3,1_{\la},0_a,1_b} 
+\sum_{\la} A_{4,\la} \ket{4,1_{\la},0_a,0_b}$ .
With the Hamiltonian (\ref{ham}), the Schr\"odinger equation leads to the 
following set of equations for the slowly-varying (during an optical cycle) 
probability amplitudes $A_j$:
\begin{mathletters}
\label{ampls}
\begin{eqnarray}
\frac{\partial A_1}{\partial t} &=& i \Om_a^* A_2 
, \label{ampl1} \\
\frac{\partial A_2}{\partial t} &=& [i \De_a - \ga_2] A_2 +
i \Om_a A_1 + i \int d \om_{\la} \rho(\om_{\la}) g_{\la} A_{3,\la} 
, \label{ampl2} \\
\frac{\partial A_{3,\la}}{\partial t} &=& 
[i (\De_a - \De_{\la}) - \ga_3] A_{3,\la} +
i g_{\la}^* A_2 + i \Om_b^* A_{4,\la} , \label{ampl3} \\
\frac{\partial A_{4,\la}}{\partial t} &=& 
[ i (\De_a - \De_{\la} + \De_b) - \ga_4] A_{4,\la} + 
i \Om_b A_{3,\la} , \label{ampl4}
\end{eqnarray}
\end{mathletters}
where $\Om_i = \mu_{jk} \bra{0_i} \hat{E}_i \ket{1_i} /\hbar = g_i $ 
is the Rabi frequency of the corresponding field at the position
of the $l$th atom, $\De_a = \om_a - \om_{21}$, $\De_b = \om_b - \om_{43}$ and 
$\De_{\la} = \om_{\la} - \om_{23}$ are the detunings from the respective 
atomic transition frequencies, and $\ga_j$ ($j=2,3,4$) is the relaxation rate 
of level $\ket{j}$, which accounts for both spontaneous radiative decay 
at a frequency far from PBG so that it is treated as a Markovian process
(Wigner-Weisskopf approximation) and nonradiative (mainly vibrational) 
relaxation in the PC. 

The Wigner-Weisskopf approximation is inapplicable near the PBG where the DOM 
varies rapidly \cite{kofman}. Thus, in the vicinity of the transition frequency
$\om_{23}$, we must exactly integrate the last term of Eq. (\ref{ampl2}) for 
the specific PBG model employed. To solve Eqs. (\ref{ampls}), we make the 
{\it weak-field approximation} $A_1 \simeq 1$ (much less than one photon per 
atom) and use the second-order perturbation theory thus obtaining the 
steady-state expressions for the atomic response. In doing so, one can see 
that the probability amplitude $A_{4,\la}$ is inversely proportional to 
the detuning $\De_b$. Then, near the Raman resonance $\De_a \simeq \De_{\la}$, 
the right-hand side of Eq. (\ref{ampl3}) contains a term 
$\ga_4 |\Om_b|^2/(\De_b^2 + \ga_4^2)$ 
resulting in an additional relaxation of the amplitude $A_{3,\la}$ and thus
destroying the coherence between levels $\ket{1}$ and $\ket{3}$ which induces 
the absorption of the $\om_a$ photon. To minimize this decoherence, we take 
$|\De_b| \gg |\Om_b|, \ga_4$, so that the depletion of $\Om_b$ can safely be 
neglected. In the opposite case of small detuning $|\De_b| \leq \ga_4$, 
it has been suggested to use this effect for constructing a sensitive 
two-photon switch \cite{harris}.

Under these conditions, one obtains effectively free-space propagation of the 
$\hat{E}_b$ field. By contrast, the Rabi frequency of the $\hat{E}_a$ field, 
in the slowly varying envelope approximation, obeys the following propagation 
equation 
\begin{equation}
\left[\frac{\partial }{\partial z} + 
\frac{1}{v_g} \frac{\partial }{\partial t}\right] \Om_a = 
i \alpha \Om_a \; , \label{maxw}
\end{equation}
with the solution $\Om_a(z,t) = \Om_a(0,t-z/v_g) \exp(i \alpha z)$. Here the 
{\it macroscopic} complex polarizability $\alpha$ is given, under the 
weak-field linear-response assumption, by
\begin{equation}
\alpha = \alpha_0 \frac{ i \ga_2}{\ga_2 - i \De_a + I} \; , \label{alpha}
\end{equation}
where 
\[
\alpha_0 = \frac{|\mu_{12}|^2 \om_{21} N}{2 \eps_0 c \hbar n_a \ga_2} 
\equiv \sigma_0 N
\]
is the linear resonant absorption coefficient on the atomic transition 
$\ket{1} \to \ket{2}$, with $\sigma_0$ the resonant absorption cross-section, 
$N$ the density of doping atoms and $n_a$ the (averaged) refraction index for 
the $\om_a$ photon, and  
\begin{equation}
I = \int \frac{d \om_{\la} \rho (\om_{\la}) |g_{\la}|^2}
{\ga_3 - i (\De_a - \De_{\la})  + 
|\Om_b|^2 [\ga_4 - i (\De_a - \De_{\la} + \De_b)]^{-1}} \;  \label{int}
\end{equation}
is the integral of the saturation factor over the structured DOM.
The group velocity $v_g$ is expressed as
\begin{equation}
v_g = \frac{\partial \om_a}{\partial k_a} = \left[ \frac{n_a}{c} 
+ \frac{\partial \text{Re}(\alpha)}{\partial \om_a} \right]^{-1} 
\; . \label{v_group}
\end{equation}
At the exit from the medium $z = \zeta$, the delay time $T_{\text{del}}$ 
of the field, relative to the passage time $T_0 = \zeta n_a / c$ through a 
passive medium, can be written as 
\begin{equation}
T_{\text{del}} =\frac{\partial \text{Re}(\alpha)}{\partial \om_a} \zeta \;  
. \label{delay}
\end{equation}

The slowly varying field-propagation equation (\ref{maxw}) must be justified
upon examining the group-velocity dispersion
\begin{equation}
D = \frac{\partial^2 k_a}{\partial \om_a^2} = 
\frac{\partial^2 \text{Re}(\alpha)}{\partial \om_a^2} \; , \label{gvd}
\end{equation}
which is responsible for the spreading and reshaping of the photon pulse.
In analogy with the weakly interacting Bose gas, $D$ can be interpreted as
being inversely proportional to the photon ``mass''. In what follows, 
conditions such that $D$ is small are discussed for $\om_a$. Due to the 
weak-field approximation $A_1 \simeq 1$ and large detuning  $\De_b$, the 
$\om_b$ photon propagation is nearly free and its group velocity dispersion 
is negligible.

To calculate integral (\ref{int}), we assume the isotropic PBG model 
\cite{kofman,john_wang,john_quang,pl} with the atoms doped at the positions
of the local defects in the PC separated by a distance $d$ from each other.
These ``impurities'' of the crystal structure form the defect modes in the PBG,
which are localized at each atomic site in a volume $V_d \simeq (r L)^3$ of 
several $(r)^3$ lattice cells $L^3$ \cite{joannopoulos,defect,kur}. In the 
dilute regime $d>rL$, one can neglect dipole-dipole interactions and tunneling 
(``hopping'') of photons  between the atoms \cite{john_wang,rddi} and the 
defect modes can serve as high-Q cavities. For $\om_{23}$ near (or within) 
the PBG frequency, we have $L \simeq \pi c /\om_{23}$. Hence, the dilute regime
limits the dopant density to $N<(\om_{23}/\pi c r )^3$. Then, in the vicinity 
of the upper edge $\om_e$ of the PBG, the DOM function can be written as
\begin{equation} 
\rho (\om)  = \rho_d \de (\om - \om_d) + 
\rho_e \frac{\Theta (\om - \om_e)}{\sqrt{\om - \om_e}} \; , \label{rho}
\end{equation}
where $\Theta (\om)$ is the Heaviside step function, $\rho_d$ and $\rho_e $ are
PC-specific constants \cite{kofman,pl}, and $\om_d$ is the frequency of the 
defect mode. The integration of Eq. (\ref{int}), with $\rho(\om_{\la})$ given 
by Eq. (\ref{rho}) and $\De_b \gg \De_{a,\la}$, leads to
\begin{equation}
I = \frac{\beta_d^2}{\ga_{31} - i(\De_a -\De_d - s_3)} -
\frac{\beta_e^{3/2}}{\sqrt{i \ga_{31} + (\De_a -\De_e - s_3)}} 
\; , \label{int_res}
\end{equation}
where $\De_{d,e} = \om_{d,e} - \om_{23} \ll \om_{23}$ are the detunings of the
defect-mode and PBG-edge frequencies from the atomic resonance $\om_{23}$, 
$\ga_{31} = \ga_3 + \ga_4 |\Om_b|^2 /\De_b^2$ the 
$\ket{1}\leftrightarrow\ket{3}$ decoherence rate, $s_3 = |\Om_b|^2/\De_b$ 
the ac Stark shift of level $\ket{3}$, and 
\cite{kofman,john_wang,john_quang,pl}
\[\beta_d^2 = \frac{|\mu_{23}|^2 \om_{23}^4}{2 \eps_0 \hbar (\pi c r)^3} 
\; , \; \; \; 
\beta_e^{3/2} = \frac{|\mu_{23}|^2 \om_{23}^{7/2}}{6 \eps_0 \hbar \pi c^3}
\]
are the coupling constants of the atom with the structured reservoir, whose 
main contributions are near $\om_d$ and $\om_e$.

\section{Results}
\label{results}

To illustrate the results of the foregoing analysis, we first plot in 
Fig. \ref{graph_comb}  the polarizability (\ref{alpha}), delay time 
(\ref{delay}) and group-velocity dispersion (\ref{gvd}) for the case of one 
incident photon $\om_a$ ($\Om_b = 0$). Clearly, {\it two frequency regions, 
$\De_a \sim \De_d$ and $\De_a \sim \De_e$}, where the absorption vanishes and, 
at the same time, the dispersion slope is steep, are of particular interest. 
One can see in Fig. \ref{graph_comb}(a),(b), where we plot the spectrum for 
two different values of the coupling constants $\beta_d$ and $\beta_e$, that 
there is, however, a substantial difference between the spectra in the 
foregoing frequency regions, for the following physical reasons:

\begin{figure}[htb]
\centerline{\psfig{figure=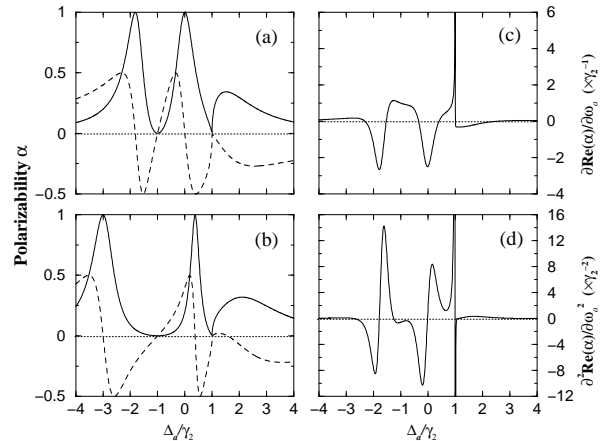,width=8cm}}
\caption{(a),(b) Imaginary (solid lines) and real (dashed lines) part of the 
complex polarizability $\alpha$, as a function of the detuning $\De_a$ for the
case $\Om_b =0$, $\alpha_0 = 1$ cm$^{-1}$, $\De_d = -1$, $\De_e = 1$ and 
$\ga_{31} = 0.001$. (a) $\beta_d = \beta_e = 1$; (b) $\beta_d = \beta_e = 2$.
(c) Delay time $T_{\text{del}}$ (per unit $z$), and (d) group velocity
dispersion coefficient $D$ as a function of the detuning $\De_a$
for the same parameters as in (a). All parameters are in units of $\ga_2$.}
\label{graph_comb}
\end{figure}

{\it a}) In the vicinity of $\De_d$, the radiation emitted by the atom at the 
frequency $\om_d$ remains confined in the defect mode for a long time as in 
a high-Q cavity. As the $\om_a$ photon wave packet approaches the $l$th atom, 
$\Om_a / \beta_d$($\ll 1$) fraction of its amplitude is transferred into the 
defect mode, inducing the corresponding population of level $\ket{3}$ 
(assuming adiabatic Raman transfer \cite{polar}). At the end of the wavepacket,
when $\Om_a \to 0$, all the population of level $\ket{3}$ returns to $\ket{1}$
and the radiation that has been confined to the defect mode is added to the 
tail of the propagating wave packet, until it encounters the next atom. 
This strong interaction of the atom with the defect mode at 
$\De_a \simeq \De_d$ splits the spectrum by the amount equal roughly to 
$2\beta_d$ and causes EIT \cite{eit}. The transparency window is rather broad 
and is given by the inverse Lorentzian [see the first term on the right-hand 
side of Eq. (\ref{int_res})]. The corresponding group velocity is much smaller 
than the speed of light:
\begin{equation}
v_g \simeq 
\left[\frac{\partial \text{Re}(\alpha)}{\partial \om_a} \right]^{-1} 
\sim \frac{\beta_d^2 }{\ga_2 \alpha_0} \ll c \; , \label{v_gr}
\end{equation}
which leads to a large delay time $T_{\text{del}} = \zeta/ v_g$ 
[Fig. \ref{graph_comb}(c)]. One has to keep in mind, however, that the 
absorption-free propagation time is limited by the EIT decoherence 
time $T_{\text{del}} < \ga_{31}^{-1}$ \cite{eit}, which imposes a limitation 
on the length $\zeta$ of the active PC medium. The corresponding 
group-velocity dispersion (\ref{gvd}) is small (a large photon mass), and, 
therefore, will not cause much spread of the $\om_a$ wave packet 
[Fig. \ref{graph_comb}(d)].  

{\it b}) In the vicinity of $\De_e$, the strong interaction of the atom with 
the continuum near the band edge $\om_e$ causes the Autler-Townes splitting 
of level $\ket{2}$ into a doublet with a separation equal roughly to $\beta_e$.
One component of this doublet is shifted out of the PBG while the other one 
remains within the gap and forms the so called photon-atom bound state 
\cite{kofman,john_wang,john_quang,pl,knight}. Consequently, there is vanishing 
absorption and rapid variation of the dispersion at $\De_a \simeq \De_e$. 
Here the delay time can be large, but the group velocity dispersion is also 
very large [Fig. \ref{graph_comb}(c),(d)]. Hence, as $\om_a$ is tuned very 
close to the band edge, the transmitted pulse shape is distorted. 
Since the transparency region is very narrow with a width 
$\De \om \sim \ga_{3}$($\ll \ga_2,\beta_e$), for an absorption-free 
propagation of the $\om_a$ photon, the temporal width of its wavepacket 
$\tau_a = l_a/c$ should satisfy the condition $\tau_a > \pi /\De \om$. 
Simultaneously, a small deviation from the condition $\De_a = \De_e$ will 
lead to a {\it strong increase} in the absorption of the $E_a$ field.   

\begin{figure}[htb]
\centerline{\psfig{figure=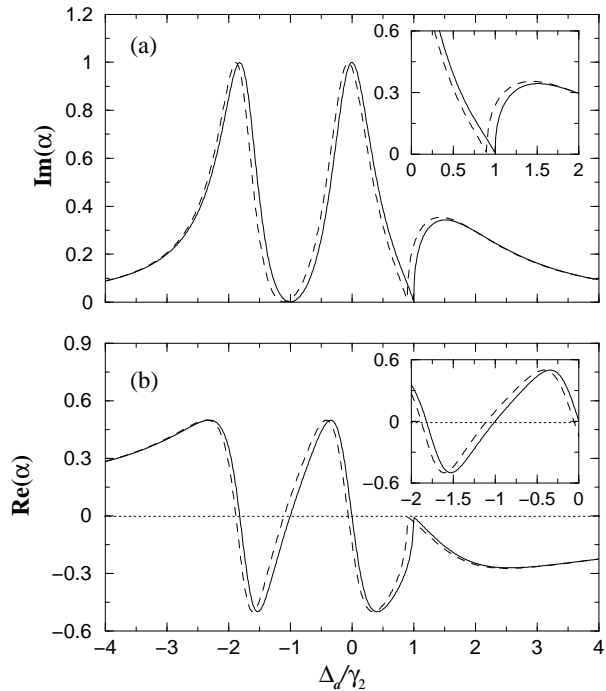,width=8cm}}
\caption{(a) Imaginary, and (b) real part of the complex polarizability 
$\alpha$ as a function of the detuning $\De_a$ for the case $\Om_b =0$ 
(solid lines) and $\Om_b =g_b$ (dashed lines). All parameters are the same 
as in Fig. \protect\ref{graph_comb}(a) and $s_3 = - 0.1$ 
(i.e., $\De_b = -10 |\Om_b|^2$). The insets magnify the important frequency
regions.}
\label{graph_s}
\end{figure}

Let us now switch on the $\om_b$ photon. As seen from Eq. (\ref{int_res}), 
its effect is merely to shift the spectrum by the amount equal to 
$s_3$ (Fig. \ref{graph_s}). This shift, however, will have different 
implications in the two frequency regions distinguished above:

i) If $s_3 \ll \beta_d$, i.e., the Stark shift is smaller than the width of 
the transparency window at $\De_d$, the medium will still remain transparent 
for a $\om_a$ photon with the detuning $\De_a = \De_d$, but its phase will 
experience an {\it appreciable nonlinear shift} $\phi_a$ given by 
\begin{equation}
\phi_a = \text{Re}(\alpha) z \simeq
- \frac{\partial \text{Re}(\alpha)}{\partial \om_a} s_3 z \sim 
- \frac{\ga_2 \alpha_0}{\beta_d^2} s_3 z \; . \label{ph_sh_add}
\end{equation}

ii) On the other hand, for a $\om_a$ photon with the detuning $\De_a = \De_e$,
the medium, which is transparent for $\Om_b = 0$, having
\begin{equation}
\text{Im}(\alpha)\simeq \frac{\ga_2 \alpha_0}{\beta_e^{3/2}}
\sqrt{\frac{\ga_{3}}{2}} \ll \alpha_0 \; , \label{im_alpha}
\end{equation}
will become highly absorptive (opaque) even for such a small frequency shift 
as $s_3$ (provided $|s_3|>\De \om$), changing (\ref{im_alpha}) to
\begin{equation}
\text{Im}(\alpha)\simeq \left\{ \begin{array}{lc}
\frac{\ga_2^2 \alpha_0}{\beta_e^3} s_3 , & s_3 > 0 \\
\frac{\ga_2 \alpha_0}{\beta_e^{3/2}} \sqrt{|s_3|} , & s_3 < 0 
\end{array}
\right. \; , \label{im_alpha_s3}
\end{equation}
and thus acting as an {\it ultrasensitive, effective switch}. 
These are the main results of the present work.

The remaining question is how to maximize the interaction between the $\om_a$ 
photon, which propagates with a small group velocity (\ref{v_gr}), and the 
$\om_b$ photon, which propagates with a velocity close to the speed of light. 
A possible technique to achieve equal group velocities for both photons, as 
suggested in Ref. \cite{lukin}, is to have a second kind of $\Lambda$-atoms 
in the interaction region and apply a driving field that would result in EIT 
and reduced group velocity for the $\om_b$ photon. To keep our scheme simple, 
we have chosen here not to adopt a similar approach. 

In our system, the interaction between the photons is maximized if: 
they enter the medium simultaneously; the transverse shapes of their tightly
focused ($\sigma_{a,b} \sim \sigma_0$) wavepackets overlap completely; 
and the quantization (wavepacket) length $\l_b$ of the $\om_b$ photon 
satisfies the condition $(\l_b + \zeta)/c \leq \zeta/v_g$. 
Then the $\om_b$ photon leaves the medium not later than the $\om_a$ photon. 
The effective interaction length between the two photons is therefore
$z_{\text{eff}} \sim l_b v_g/c \leq \zeta$, after which the two wavepackets 
slip apart. This effective interaction length has the following implications:

1) Since $s_3 \propto 1/l_b$, the interaction-induced phase shift 
(\ref{ph_sh_add}) saturates with distance and is the same for all $l_b$ 
satisfying the above condition:
\begin{equation}
\phi_a = - \frac{\partial \text{Re}(\alpha)}{\partial \om_a} s_3 
\frac{l_b v_g}{c} \simeq
- \frac{|\mu_{34}|^2 \om_b}{2 \eps_0 \hbar c \sigma_b \De_b} \; . \label{ph-sh}
\end{equation}

2) As was mentioned in Sec. \ref{theory} and carefully analyzed in 
Ref. \cite{harris}, in the case of a small detuning $|\De_b| \leq \ga_4$, the 
presence of the $\om_b$ photon will result in the destruction of EIT in the 
vicinity of $\De_a = \De_d$ and, consequently, the strong absorption, which is 
determined by
\[
\text{Im}(\alpha) \simeq \frac{\ga_2 \alpha_0 |\Om_b|^2}{ \ga_4 \beta_d^2 }  .
\]
With the effective interaction length $z_{\text{eff}} \sim l_b v_g/c$, where 
$v_g$ is given by Eq. (\ref{v_gr}), the power loss at the exit from the PC is 
\begin{equation}
2 \text{Im}(\alpha) z_{\text{eff}} \simeq
\frac{|\mu_{34}|^2 \om_b}{\eps_0 \hbar c \sigma_b \ga_4} \; . \label{abs}
\end{equation}
Thus, in the vicinity of the two-photon (Raman) resonance with the defect mode,
$\De_a = \De_d$, the absorption (\ref{abs}) ($|\De_b| \leq \ga_4$), as well as 
the phase shift (\ref{ph-sh}) ($|\De_b| \gg |\Om_b|,\ga_4$), {\it saturate} 
over a distance equal to $z_{\text{eff}}$. 

3) The dramatic advantage of the present scheme over conventional EIT 
schemes \cite{imam,harris} is that, in the vicinity of the two-photon 
resonance with the band edge frequency, $\De_a = \De_e$, the group velocity 
of the $\om_a$ photon wavepacket is close to the speed of light in the 
presence of the nearly-free propagating $\om_b$ photon wavepacket 
[see Fig. \ref{graph_comb}(c)]. Therefore, there is no velocity mismatch of
the two photons and {\it no saturation} of the absorption with distance, 
the interaction length being close to the length $\zeta$ of the PC. 
Thus the absorption probability of the $\om_a$ photon, 
$1-\exp[-{2 \text{Im}(\alpha) \zeta}]$, where $\text{Im}(\alpha)$ is given by
(\ref{im_alpha_s3}), can be made {\it arbitrary close to unity} by choosing 
a long enough PC. 

It can be checked that, for the parameter values used in Fig. \ref{graph_s},
and $\mu_{34} \sim 1$ a.u., $\om_b = 4\times 10^{15}$ rad/s 
($\lambda_b = 470$ nm), $|\De_b| = 10 \ga_2$, 
$\ga_2 \sim 5 \times 10^7$ s$^{-1}$, $\sigma_b \sim 10^{-10}$ cm$^{-2}$,
we obtain the power loss $2 \text{Im}(\alpha) \zeta \sim 0.46 \zeta$ at 
$\De_a = \De_e$ and the phase shift $\phi_a \sim 0.1$ rad at $\De_a = \De_d$. 
Thus, the presence of one $\om_b$ photon induces either strong absorption or 
a large phase shift of the $\om_a$ photon, depending on the frequency region 
employed.

\section{Discussion}
\label{discuss}

The foregoing results have demonstrated the ultrahigh sensitivity of photonic
absorption or phase shift in a doped PC to the presence of an additional
photon. Using the procedure of Ref. \cite{harris} one can employ these 
features to construct two types of entanglement between the photons $\om_a$ 
and $\om_b$: 

{\it a}) Suppose that a photon pair $\ket{1}_{\om_a} \ket{1}_{\om_b}$ is 
simultaneously generated in some parametric or other process. The $\om_a$ 
photon enters the doped PC. The $\om_b$ photon is split between the two arms 
of a 50\%-50\% beam splitter, resulting in an entangled state 
$1/\sqrt{2}(\ket{10}_{\om_b} + \ket{01}_{\om_b})$. One arm of this state, say 
$\ket{10}_{\om_b}$, enters the PC together with the $\om_a$ photon, while the 
other arm, $\ket{01}_{\om_b}$, does not. In the case of the $\om_a$ photon 
with the detuning $\De_a \simeq \De_e $, after passing through the PC, the 
resulting state of the system has  the fully entangled form
\begin{eqnarray}
& & \frac{1}{\sqrt{2}} (\ket{10}_{\om_b} + \ket{01}_{\om_b}) \otimes 
\ket{1}_{\om_a} \nonumber \\  & &
\to \frac{1}{\sqrt{2}} ( \ket{10}_{\om_b} \ket{0}_{\om_a} +
\ket{01}_{\om_b} \ket{1}_{\om_a}) \; ,
\end{eqnarray}
wherein the states in which the $\om_a$ photon is or is not absorbed
are equally superposed.

{\it b}) In the case of the $\om_a$ photon being initially in a state with a 
rather well defined phase (i.e., coherent state) $\ket{\alpha}$ and 
having the detuning $\De_a \simeq \De_d $, after passing through the PC, the 
resulting state of the system is given by
\begin{eqnarray}
& & \frac{1}{\sqrt{2}} (\ket{10}_{\om_b} + \ket{01}_{\om_b}) \otimes 
\ket{\alpha}_{\om_a} \nonumber \\  & &
\to \frac{1}{\sqrt{2}} ( \ket{10}_{\om_b} \ket{e^{i \phi_a} \alpha}_{\om_a} +
\ket{01}_{\om_b} \ket{\alpha}_{\om_a}) \; , \label{ph_ent}
\end{eqnarray}
wherein the states in which the $\om_a$ photon does or does not acquire the 
phase shift $\phi_a$ are equally superposed. The state (\ref{ph_ent}) is
fully entangled only if $\ket{e^{i \phi_a} \alpha}$ and $\ket{\alpha}$ do
not overlap in the phase plane (the same requirement as for a ``Schr\"odinger
cat'' state in a PC \cite{kur}).

These features can be used to appreciably advance towards the goal of 
producing entangled states of radiation or logical photon switches for 
quantum information processing, owing to the unique advantages of the doped
PCs over conventional EIT schemes \cite{imam,harris,lukin} or high-Q cavities
\cite{cavity}: (i) No control lasers are needed to create the quantum 
interference responsible for the EIT effects. (ii) The system parameters can 
easily be adjusted to provide full two-photon entanglement via absorption. 
(iii) Several doped PCs can be combined via dispersive couplers either to
enhance the accumulated nonlinear phase-shift or to perform a chain of
logical operations. The fact that the band structure of dilutely doped 
PCs is {\it insensitive} to the locations or exact concentration of the 
dopants and defects should facilitate the incorporation of several PCs
in one network.

\acknowledgments

This work was supported by the US--Israel BSF and the Feinberg 
Fellowship (D.P.).

\end{document}